\documentstyle[12pt,epsf]{article}
\begin{document}
%%%%%%%%%%%%%%%%%%%%%%%%%%%%%%%%%
%%%%%%%%%%%%%%%\def\baselinestretch{1.5}
\catcode`@=11
\def\marginnote#1{}
%%%%%%%%%%%%%%%%%%%%%%%%%%%%%%%%%%%%%%%%%
%
%%%%%%%%%%%%%%%%%% begin definitions %%%%%%%%%%%%%%%%%%
\newcommand{\newc}{\newcommand}
\newc{\be}{\begin{equation}}
\newc{\ee}{\end{equation}}
\newc{\bea}{\begin{eqnarray}}
\newc{\eea}{\end{eqnarray}}
\newc{\ie}{{\it i.e.}}
\newc{\eg}{{\it eg.}}
\newc{\etc}{{\it etc.}}
\newc{\etal}{{\it et al.}}

\newc{\ep}{e^+}
\newc{\epem}{e^+e^-}
\newc{\mz}{m_Z}
\newc{\mispt}{{{\not\! p}_T}}

\newc{\gsim}{\lower.7ex\hbox{$\;\stackrel{\textstyle>}{\sim}\;$}}
\newc{\lsim}{\lower.7ex\hbox{$\;\stackrel{\textstyle<}{\sim}\;$}}

\newc\rcons{R\mbox{$\surd$}}	\newc\rviol{R\mbox{$\large\times$}}

\newc{\hc}{{\it h.c.}}
\newc{\lam}{\lambda}
\newc{\lampr}{\lam^\prime}
\newc{\lamdp}{\lam^{\prime\prime}}
%
% superfields:
\newc{\superhu}{\hat H_u}	\newc{\superhd}{\hat H_d}
\newc{\superl}{\hat L}		\newc{\superr}{\hat R}
\newc{\superec}{\hat {e}^c}
\newc{\superq}{\hat Q}
\newc{\superu}{\hat U}
\newc{\superd}{\hat D}
\newc{\superuc}{\hat {u}^c}
\newc{\superdc}{\hat {d}^c}
%
% superpartners (bosons):
\newc{\tildef}{\widetilde f}

\newc{\tildel}{\widetilde L}	\newc{\tilder}{\widetilde R}
\newc{\tildenu}[1]{\widetilde{\nu}_{#1}}
\newc{\tildee}[1]{\widetilde e_{#1}}
\newc{\tildeec}[1]{\widetilde{e}^c_{#1}}
\newc{\tildeel}[1]{{\widetilde e}_{L#1}}
\newc{\tildeer}[1]{{\widetilde e}_{R#1}}
\newc{\tildeq}{\widetilde Q}
\newc{\tildeu}[1]{\widetilde u_{#1}}
\newc{\tildeuc}[1]{\widetilde{u}^c_{#1}}
\newc{\tildeul}[1]{{\widetilde u}_{L#1}}
\newc{\tildeur}[1]{{\widetilde u}_{R#1}}
\newc{\tilded}[1]{\widetilde d_{#1}}
\newc{\tildedc}[1]{\widetilde{d}^c_{#1}}
\newc{\tildedl}[1]{{\widetilde d}_{L#1}}
\newc{\tildedr}[1]{{\widetilde d}_{R#1}}
				\newc{\nubar}[1]{{\bar\nu}_{#1}}
				\newc{\ebar}[1]{{\bar e}_{#1}}
\newc{\ec}[1]{e^c_{#1}}		\newc{\ecbar}[1]{{\bar e}^c_{#1}}
				\newc{\ubar}[1]{{\bar u}_{#1}}
\newc{\uc}[1]{u^c_{#1}}		\newc{\ucbar}[1]{{\bar u}^c_{#1}}
				\newc{\dbar}[1]{{\bar d}_{#1}}
\newc{\dc}[1]{d^c_{#1}}		\newc{\dcbar}[1]{{\bar d}^c_{#1}}
\newc{\pl}{P_L}		\newc{\pr}{P_R}
\newc{\mw}{m_W}
\newc{\tw}{\tan\theta_w}
\newc{\tanb}{\tan\beta}
\newc{\sinb}{\sin\beta}
\newc{\cosb}{\cos\beta}

\newc{\ra}{\rightarrow}

\newc{\lsp}{{\chi}}
\newc{\mchi}{m_{\chi}}	\newc{\sgnmchi}{{\epsilon_\chi}}
\newc{\snumu}{\widetilde\nu_\mu}
\newc{\snutau}{\widetilde\nu_\tau}
\newc{\slepton}{\widetilde l}
\newc{\squark}{\widetilde q}
\newc{\vev}{{\it vev}}
\newc{\VEV}[1]{\langle #1 \rangle}
\newc{\tauchi}{\tau_\chi}
\newc{\fprime}{f^\prime}   \newc{\fdprime}{f^{\prime\prime}}
%%%%%%%%%%%%%%%%%% end of definitions %%%%%%%%%%%%%%%%%%
%%%%%%%%%%%%% end of defs %%%%%%%%%%%%%%%
%\begin{titlepage}
\begin{flushright}
%\hfill
UM-TH-93-21\\
%\hfill
hep-ph/9309208\\
%\hfill
August 1993\\
\end{flushright}
\vskip 0.6in
\begin{center}
{\large \bf ON R-PARITY VIOLATION AT $e^+ e^-$ COLLIDERS
\footnote{\tenrm\baselineskip=11pt Invited talk at the {\em Workshop  
on Physics
and Experiments with Linear $e^+e^-$ Colliders}, Waikaloa, Hawaii,  
April
26-30, 1993, to appear in the Proceedings.}
\\}
\vskip .4in
{\large LESZEK ROSZKOWSKI}
\vskip .1in
{\em leszek@leszek.physics.lsa.umich.edu\\
     Randall Physics Laboratory,\\
     University of Michigan,\\
     Ann Arbor, MI 48109-1129, USA}
\end{center}
\vskip .2in
\begin{abstract}
\noindent
I discuss several promising $R$-parity violating processes at $e^+  
e^-$
colliders.
\end{abstract}

%	\end{titlepage}
%
     \setlength{\baselineskip}{14pt}
\setcounter{footnote}{0}
\setcounter{page}{1}	% 1
\setcounter{section}{0}
\setcounter{subsection}{0}
%		\tableofcontents
%	\newpage
% BODY

\vskip 1cm
\section{R-Parity Violation: Pandora's Box of SUSY}

In the Minimal Supersymmetric Standard Model several new interactions
are allowed if $R$-parity is explicitly broken.
The superpotential $W=W_{\rcons} + W_{\rviol}$ now  
reads~\cite{hallsuzuki}
\be
W_{\rcons}= h^U_{ij} \superq_i\superhu\superuc_j
           + h^D_{ij} \superq_i\superhd\superdc_j
	   + h^E_{ij} \superl_i\superhd\superec_j
	   + \mu \superhd\superhu
\label{spotential-rcons}
\ee
and
\be
W_{\rviol}= \lam_{ijk} \superl_i \superl_j \superec_k +
       \lampr_{ijk} \superl_i \superq_j \superdc_k +
       \lamdp_{ijk} \superuc_i \superdc_j \superdc_k +
	 \kappa_i\superl_i\superhu.
\label{spotential-rbreak}
\ee

Note that $i,j,k=1,2,3$, $\lam_{ijk}=-\lam_{jik}$ because
$\superl_i \superl_j=\epsilon_{ab} \superl_i^a \superl_j^b=
\superl_i^1 \superl_j^2 -\superl_i^2 \superl_j^1$, and
$\lamdp_{ijk}=-\lamdp_{ikj}$ to counter-balance the antisymmetricity  
of the
color indices of  the superfields $\superdc$. In general there are  
thus 39
$L$-number ($\lam,\lampr,\kappa$) and 9 $B$-number violating
($\lamdp$) couplings.

Since the superfields $\superl_i$ ($i=1,2,3$) and $\superhd$
carry the same quantum numbers, one can rotate them by an arbitrary
$SU(4)$ rotation $U$.
This operation in general
mixes the Yukawa couplings $h^D$ with $\lampr$, $h^E$ with $\lambda$,
and $\kappa_a$ with $\mu$, respectively.
This implies that none of these terms in $W$ can be
a priori neglected since a rotation $U$ would in general re-generate
them.

In particular, following Hall and Suzuki~\cite{hallsuzuki}, one  
usually rotates
away the terms $\kappa_i\superl_i\superhu$
by suitably choosing the rotation matrix $U$.
No other terms can be eliminated without further assumptions. After  
these terms
have been rotated away, only the lepton fields can still mix among  
themselves.

It should be stressed that by rotating away the $\kappa$-terms one,
in general, induces sneutrino \vev\,s~\cite{barbierihall,lr}.
One can next conveniently rotate the lepton doublet superfields so  
that only
the
sneutrino of one generation will acquire a \vev\ and the  
corresponding neutrino
will acquire mass at the tree level. The masses of
the other two neutrinos will be induced via one-loop diagrams.

A non-zero sneutrino \vev\ will also
mix leptons with charginos and
neutrinos with neutralinos, and thus induce new couplings:
$Z^0 \nu \chi^0_i$, $Z^0 l^\pm \chi^\mp_j$, $W^\pm \nu \chi^\mp_j$,  
and
$W^\pm l^\mp \chi^0_i$ ($i=1,4$ and $j=1,2$)~\cite{barbierihall,lr}.  
While
these are
strongly suppressed by stringent constraints on lepton universality,  
they do
constitute a separate class of $R$-parity violating terms and a  
priori cannot
be ignored.

The lagrangian terms ${\cal L}= {\cal L}_\lam + {\cal L}_{\lampr} +  
{\cal
L}_{\lamdp}$ corresponding to $W_{\rviol}$ read
\begin{equation}
{\cal L}_\lam= (\lam_{ijk}-\lam_{jik})
[ \tildenu{j} \ebar{k} \pl e_i + \tildee{i} \ebar{k}\nu_j +
\tildeec{k} \ecbar{i}\nu_j ] + \hc
\label{lagrangian-lam}
\end{equation}
\begin{equation}
{\cal L}_{\lampr}= \lampr_{ijk}
[ \tildee{i} \dbar{k} \pl u_j - \tildeu{j} \dbar{k} \pl e_i +
\tildedc{k} \ucbar{j}\pl e_i
-\tildenu{i} \dbar{k} \pl d_j - \tilded{j} \dbar{k} \nu_i
-\tildedc{k} \dcbar{j}\nu_i] + \hc
\label{lagrangian-lampr}
\end{equation}
and
\begin{equation}
{\cal L}_{\lamdp}= -{1\over2}(\lamdp_{ijk}-\lamdp_{ikj})
[ \tildedc{k} \ubar{i}  \pl \dc{j} +
\tildeuc{i} \dbar{j}  \pl \dc{k} ] + \hc
\label{lagrangian-lamdp}
\end{equation}
Note that $i,j,k=1,2,3$, and the symmetry properties of the
Yukawas are now explicitly displayed.

Since $R=(-1)^{3B+2s+L}$, where $B, s, L$ are the baryon, spin, and
lepton numbers of a given (s)particle, respectively,
$R$-parity is broken if either $L$- or $B$-number is broken (or  
both).
Simultaneous violation of both $L$ and $B$ leads to a fast proton
decay. (More precisely, this is true
when simultaneously $\lampr\neq0$ and
$\lamdp\neq0$.
But one cannot simply set $\lampr_{ijk}$ to zero since
it would be next regenerated by rotating out the $\kappa$-terms.)

Below I discuss several signatures of potential interest at $\epem$
accelerators in the case of explicit $R$-parity breaking.
(It occurs that very similar phenomenology actually results also in  
the case of
spontaneous $R$-parity breaking~\cite{valle}.)
I mainly focus on $L$ violating
processes at $\epem$ machines but will
also comment on the case $\Delta B\neq0, \Delta L=0$.
The phenomenology of $R$-parity breaking at hadronic colliders has  
been
discussed, \eg, in Refs.~\cite{dreinerross,savas} and at HERA, \eg,  
in
Ref.~\cite{butterworth}.

\vskip 0.6cm
\section{Unstable LSP}

If $R$-parity is broken there is no distinction
between particles and sparticles
and the concept of the lightest supersymmetric
particle (LSP) is ill-defined. (For example, if $\Delta L\neq0$ then
the sneutrinos and neutral Higgs scalars carry the same quantum  
numbers. The
same is true with the neutrinos and neutralinos.)
Nevertheless it is still convenient to
use it in the limit $\lam, \lampr, \lamdp \ll g$
and $\langle
\tildenu{}\rangle\ll\mz$, \ie, when $R$-parity is
`almost' unbroken. In the present study I will assume that the LSP is
the lightest of the four neutralinos, $\chi\equiv\chi_1^0$  
(neglecting
their small mixing with
the neutrinos, induced by the sneutrino \vev\,s).
Other choices, like the
stop, slepton, chargino, or sneutrino are considered much more exotic
and/or constrained by cosmology but have been claimed not to be fully
excluded yet~\cite{grt}.
\begin{center}
\epsfxsize=2.75in
\hspace*{0in}
\epsffile{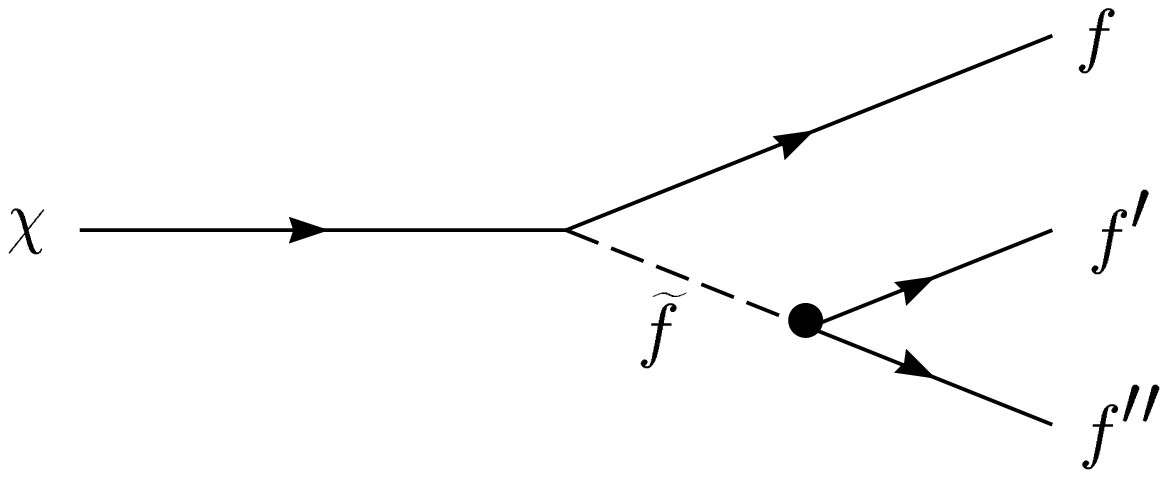}

\vspace{-1cm}
\parbox{2.5in}{\small\hfil Fig.~1. Indirect decays  of $\chi$.\hfil}
\end{center}

An unstable LSP can decay into `ordinary' matter.
There are essentially three regions of interest for the LSP lifetime
$\tauchi$. If $\tauchi\lsim 10^{-8}\, sec$, $\chi$ will decay inside  
the
detector. (For more details, see, \eg,  
Refs.~\cite{dreinerross,dawson}.)
Otherwise it escapes detection.
For $\tauchi\gsim 10^{17}\, sec$, the neutralino is also stable
cosmologically and thus constitutes a good candidate for dark
matter in the Universe. In this case $R$-parity would be effectively
conserved.
In what follows I will discuss the cases of both the LSP decaying  
inside
and stable in the detector even though cosmological arguments
suggest much larger $\tauchi$.
\begin{center}
\epsfxsize=2.75in
\hspace*{0in}
\epsffile{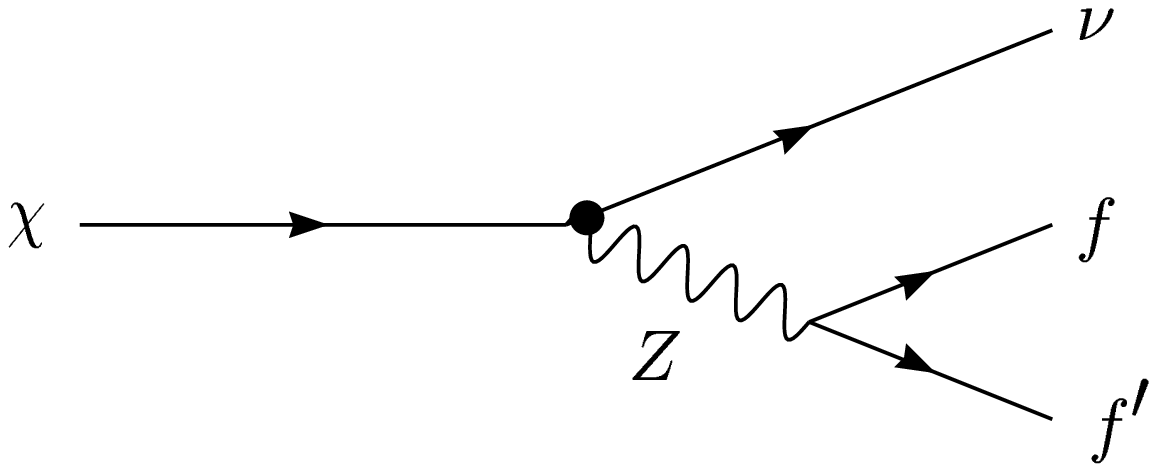}

\vspace{0.75cm}
\parbox{4.5in}{\small Fig.~2. Direct $\chi$ decay into $\nu f\bar f$.
A similar diagram leads to the final state $l f \fprime$ via the  
$W$-exchange.}
\end{center}

There are two classes of decays the LSP can undergo.
\noindent
In {\bf indirect}
decays
$\chi\ra f\tildef$ followed by $\tildef\ra f^\prime f^{\prime\prime}$
(Fig.~1), where $f$ stands for any lepton or quark field.
The LSP $\chi$ can therefore decay into $\nu_i l_j\bar l_k$, $\nu_i  
q_j\bar
q_k$, or $l_i q_j\bar q_k$ via $\Delta L\neq0$ diagrams, or into 3  
quarks via a
$\Delta B\neq0$ squark exchange.
\noindent
In {\bf direct} decays ($\Delta L\neq0$ only)
$\chi$ decays to $\nu f \fprime$ (Fig.~2) or to $l^\pm f \fprime$
via either real
or virtual $Z$ and $W$ exchange, respectively. Direct LSP decays are
possible due to the couplings $Z\chi\nu$ and $W^\pm\chi l^\mp$
induced by the sneutrino \vev\ discussed above.
(In this case also the (massive) neutrino may in principle be  
unstable;
the possibility I will put aside here.)
Below, for definiteness, I will only consider indirect LSP decays.
Of course including
direct decays would lead to several additional final states and thus
potentially interesting signatures.

I will discuss
three classes of processes
characteristic for $R$-parity violation in $\epem$ collisions.
\begin{enumerate}
\item
$R$-parity-breaking {\bf single sfermion production} is next followed
by its decay via either $\rviol$ or $\rcons$ couplings.
\item
A {\bf double sfermion production} involving either $\rviol$ or  
$\rcons$
couplings, with each sfermion next decaying
via $\rviol$ vertices.
\item
Sneutrino-\vev-induced processes involving {\bf no sfermion  
production}
but rather double {\em ino} (both neutrino and neutralino) production  
and
decay. They both in principle occur via $\rviol$ or $\rcons$  
vertices.
\end{enumerate}

Some of these processes lead to signatures identical to those  
resulting
from $\rcons$ processes. This adds to the complexity of the situation  
and shows
that in general
one cannot fully separate $\rviol$ processes
from the $\rcons$ ones.

\vskip 0.6cm
\section{Single Sfermion Production}
In $\epem$ accelerators one can only produce a single $\snumu$ or
$\snutau$ via $\lam_{121}$ and $\lam_{131}$, respectively. In  
addition,
there is an associated $\widetilde\nu_{\mu,\tau}Z$-production diagram  
involving
a $t$-channel
{\em  ino}-exchange
which is, however, $g^2$-suppressed. Similarly, there is a class
of $t$-channel
processes involving an associate $W$-charged-slepton production.

At present experimental constraints on the involved couplings are not
very stringent: $\lam_{121}\lsim0.04$ and
$\lam_{131}\lsim0.1$~\cite{bgh}. They also depend on various  
assumptions.

The sneutrino $\tildenu{i}$ ($i=\mu,\tau$) can next decay in a  
variety of ways.
The resulting signatures depend on whether or not the LSP(s)
produced at some point will escape from the detector. I will discuss  
both
possibilities separately.

\begin{enumerate}
\item
$\underline{\tildenu{\mu,\tau} \ra \nu\chi}$. Since the coupling
$\tildenu{} \nu\chi_i$ ($i=1,4$) is of the order of the weak gauge
coupling $g$ (for gaugino-type $\chi_i$), this decay mode may well be
dominant.

If the LSP $\chi$ escapes detection the final state is completely
invisible ($\ep e^-\ra s$-nothing).

The LSP $\chi$ can also decay (indirectly) in the detector
into a neutrino and either
two (charged) leptons or two quarks via a sfermion exchange (Fig.~1).  
In this
case the final state, either $l_i\bar l_j +\mispt$ ($i,j=1,2,3$)
or 2 jets and $\mispt$
will be, however, obscured by the background from $\ep e^-\ra ZZ$.  
More
study is needed to assess to what extent the different geometry will
help in extracting the signal.

\item
$\underline{\tildenu{\mu,\tau} \ra \nu\chi^\prime}$ (where
$\chi^\prime$ denotes one
of the heavier neutralinos). In this case $\chi^\prime$ will
cascade-decay into one or more LSPs and several fermions leading to  
many
possible final states.

\item
$\underline{\tildenu{\mu,\tau} \ra l^\pm_{\mu,\tau}\chi^\mp}$ with  
the chargino
next decaying to the LSP and
either $(l,\bar\nu_l)$ or $(q,\bar q^\prime)$.
Even if the LSP escapes detection, the final state signatures:  
$\epem\ra
\mu/\tau+l+\mispt$ or $\ra \mu/\tau +\, {\rm 2\ jets}+ \mispt$ should  
be
distinguishable from the background caused by $WW$-pairs.

\item
$\underline{\tildenu{\mu,\tau} \ra \slepton^{\pm}_{\mu,\tau} W^\mp}$.  
The
smuon (stau) next decays into the LSP and a $\mu$ ($\tau$) leading to
the same signature as in the previous point.
In both cases, if the LSP decays inside the detector, there will be  
an
additional pair of leptons or jets with missing $p_T$.

\item
$\underline{\tildenu{\mu,\tau} \ra \tildenu{\mu,\tau} Z}$.
The singly produced sneutrino may emit off a $Z$ before decaying  
along
one of the several possible patterns. The classes of final states are
the same as in point~1, and similarly one needs to worry about the
background from $ZZ$.

\item
$\underline{\tildenu{\mu,\tau} \ra \bar l_j l_k, d_r \bar d_s}$ via
$\lam_{ijk}$ ($j=2,3$; $i\neq j$, and $k=1,2,3$), and $\lampr_{rjs}$
($r,s=1,2,3$), respectively.
A detection of two charged leptons or jets of different flavors would  
provide a
striking signal for
$R$-parity violation. (Same-flavor final states will suffer from the
background due to neutral gauge boson exchange.)
These processes would also allow for a direct
reconstruction of the sneutrino mass but, being doubly suppressed
relative to $\rcons$ decays, are probably less likely to
be observed. (On the other hand, experimental bounds~\cite{bgh} on
$\rviol$ operators do not apply now~\cite{herbi-private}.)

\end{enumerate}

Several of the above processes have been discussed in Ref.~\cite{dl}.  
A
numerical example~\cite{dl} for $e^+e^-\ra \tildenu{} Z$ at the NLC  
shows that
the process should be clearly visible up to
the kinematic limit even for rather small values of $\lambda$.

It is clear that even if a single $\tildenu{\mu}$ or $\tildenu{\tau}$  
is
produced there will in general be a variety of possibly striking
signatures, {\em even} if the LSP does not decay in the detector.
This is also true in the case of hadronic  
colliders~\cite{dreinerross}.

\vskip 0.6cm
\section{Double Sfermion Production}

Pairs of sleptons and squarks can be produced in $e^+ e^-$ collisions
via $\rcons$ neutral gauge boson ($\gamma$ and $Z$)
$s$-channel exchange.
Pairs of $\tildee{}^+ \tildee{}^-$
can also be created through a diagram involving
a $t$-channel exchange of $\chi^0_i$ ($i=1,4$).
In addition, there exist $t$-channel $\rviol$ processes which are,  
however,
doubly suppressed and thus probably less important.

Each sfermion will next decay via either $\rcons$ or $\rviol$  
vertices into one
of several possible final states following the pattern discussed  
above in the
case of $\tildenu{\mu,\tau}$. For definiteness, let us focus on a  
(charged)
slepton. It can decay into $l\nu$ (via $\lambda$) or $q\bar q^\prime$  
(via
$\lambda^\prime$). But
it can also decay into $l\chi$ without breaking $R$. Clearly
in all these processes the main background will come from $WW$ pairs
but should be tractable by cutting on $m_W$.

If $\chi$ next escapes detection, the possible final
states are
\bea
e^+ e^- \ra{\widetilde l}^+{\widetilde l}^-&\ra& \bar l l +\mispt\\
						   &\ra& l + 2{\rm  
jets} +\mispt.
\label{chiescapes}
\eea

If $\chi$ decays in the detector, many more possibilities arise:
\bea
e^+ e^- \ra\widetilde l^+\widetilde l^-&\ra& 6l +\mispt,\\
	&\ra& (4 {\rm\ or\ }5)l + 2{\rm\ jets} +\mispt,\\
	&\ra& 4l + 4{\rm\ jets},\\
	&\ra& (2 {\rm\ or\ }3)l + 4{\rm\ jets} +\mispt,\\
	&\ra& 2l + 6{\rm\ jets} +\mispt.
\label{chidecays}
\eea

One can see that $R$-parity breaking can show up in a variety of  
ways.
In the case considered above a distinct signal is provided by  
multiple lepton
events and relatively little background. (Further discussion of some  
of these
processes and numerical examples in the case of the NLC
can be found
in Ref.~\cite{dl}.)

\vskip 0.6cm
\section{Ino Production}

Another class of processes involves two {\em ino}\,s in the final  
state.
Usually one considers only$\rcons$ $\chi$-pair-production processes.  
(See,
\eg, Ref.~\cite{grt}.) In addition, however, the  
sneutrino-\vev-induced
$\rviol$ couplings $Z\nu\chi^0$ and $Wl\chi$ allow for $\nu\chi$ in  
the final
state. These processes are presumably suppressed
by constraints from lepton universality but in general should not be
neglected.

\vskip 0.6cm
\section{Conclusions}
$R$-parity breaking can truly be called Pandora's box of  
supersymmetry.
Once it is allowed it leads to an enormous number of new
processes and final states. In this brief review I have
attempted to at least systematize major $R$-breaking processes in
$\epem$ collisions. Some specific numerical examples were quoted for  
the
NLC but the classification presented here is (with some  
modifications,
like $ZZ$ or $WW$ background) applies also to LEP. Clearly, an
extensive study is needed to assess all the possibilities. It is not
even easy to clearly distinguish dominant and sub-dominant processes  
without
making {\it ad hoc} assumptions regarding the couplings. Moreover, in
general one {\em cannot} separate $R$-parity violating processes from  
the
conserving ones.

On the other hand, a discovery of $R$-parity breaking could have very
profound consequences for our understanding of several fundamental
issues. Among them are, for example, hints for specific ways of GUT
symmetry breaking and the nature of dark matter in the Universe. The
task of searching for $R$-parity violation may not be an easy one but
it may well be worth the effort.

\vskip 0.6cm
\section*{Acknowledgments}

\noindent
I would like to thank H.~Dreiner and X.~Tata for comments on the  
manuscript.
\vskip 0.6cm
%		\bigskip
%		\newpage
%%%%%%%%%%%%% begin refs %%%%%%%%%%%%%%%%

%%%%%%%%%%%%% end of refs %%%%%%%%%%%%%%%

\begin{thebibliography}{99}
%
\bibitem{hallsuzuki}
L.~Hall and M.~Suzuki, {\it Nucl. Phys.} {\bf B231} (1984) 419.
%
\bibitem{barbierihall}
R.~Barbieri and L.~Hall, {\it Phys. Lett.} {\bf B238} (1990) 86.
%
\bibitem{lr}
L.~Roszkowski, in preparation.
%
\bibitem{valle}
See, \eg,
M.~Gonzalez-Garcia, J.~Romao, and J.~Valle, {\it Nucl. Phys.}
{\bf B391} (1993) 100, and references therein.
%
\bibitem{dreinerross}
H.~Dreiner and G.~Ross, {\it Nucl. Phys.} {\bf B365} (1991) 597.
%
\bibitem{savas}
S.~Dimopoulos, R.~Esmailzedeh, L.~Hall, J.-P.~Merlo, and G.~Starkman,  
{\it
Phys. Rev.} {\bf D41} (1990) 2099.
%
\bibitem{dawson}
S.~Dawson, {\it Nucl. Phys.} {\bf B261} (1985) 297.
%
\bibitem{butterworth}
J.~Butterworth and H.~Dreiner, {\it Nucl. Phys.} {\bf B397} (1993) 3.
%
\bibitem{bgh}
V.~Barger, G.~Giudice, and T.~Han, {\it Phys. Rev.} {\bf D40} (1989)  
2987.
%
\bibitem{herbi-private} H.~Dreiner, private communication.
%
\bibitem{dl}
H.~Dreiner and S.~Lola, in {\it Proceedings of the Workshop - Munich,
Annecy, Hamburg, Feb.~4 - Sep.~3, 1991}, ed. P.M.~Zerwas.
%
\bibitem{grt}
R.~Godbole, P.~Roy, and X.~Tata, CERN-TH-6613-92 (August 1992).
\end{thebibliography}
\end{document}